\documentclass[aps,showpacs,twocolumn,superscriptaddress,prb,floatfix,10pt, longbibliography]{revtex4-2}
\pdfoutput=1
\usepackage[utf8]{inputenc}
\usepackage{amsfonts}
\usepackage{amsmath}
\usepackage{amssymb}
\usepackage{bbm}
\usepackage{amsthm}
\usepackage{empheq}
\usepackage{enumitem}
\usepackage{verbatim}
\usepackage{color}
\usepackage{subfigure}
\usepackage[breaklinks]{hyperref}
\usepackage{graphicx}
\usepackage{times}

\begin{document}

\preprint{APS/123-QED}

\title{Transverse spin photocurrents in ultrathin topological insulator films}

\author{S. Movafagh}
\affiliation{Department of Physics and Astronomy,\\George Mason University, Fairfax, VA 22030, USA}
\author{P. Nikoli\'{c}}%
\affiliation{Department of Physics and Astronomy,\\George Mason University, Fairfax, VA 22030, USA}

\date{\today}

\begin{abstract}
Nonlinear helicity-dependent photocurrents have been reported in 3D topological materials lacking inversion symmetry. Here we study theoretically the charge and spin photocurrents generated by linear and circularly polarized radiation in ultrathin topological insulator films. Using time-dependent perturbation theory and detailed balance equations, we find that helical transverse spin currents are generated when the symmetry between the top and bottom film surfaces is disturbed. Such spin currents are invariant under in-plane mirror transformations, but have $s$-wave and $d$-wave  components in regard to transformations under $\pi/2$ rotations. Spin current photo-transistors and amplifiers can be based on these effects.
\end{abstract}

\maketitle


\section{Introduction}\label{sec:level1}

Topological insulators (TI) have been of great interest for their gapless modes at the crystal boundaries, protected by the time-reversal symmetry. \cite{Murakami2004, Kane2005, Kane2005a, Bernevig2006, Qi2006, Konig2007, Roy2009, yan2012topological, Hasan2011, Fu2007, Moore2007, Zhang2010, bansil2016colloquium}. They also have potential applications in spintronics \cite{wang2015spintronics, hou2020progress, weber20242024}.  Transport via spin-polarization is an attractive alternative to charge transfer due to a reduced dissipation and the ease of tuning \cite{wang2015spintronics}. A particular studied transport mechanism in topological insulators is the induction of directed spin-polarized charge currents by infrared or terahertz radiation, also known as photogalvanic effect (PGE) \cite{junck2013photocurrent, plank2018infrared, takeno2018optical, kastl2015ultrafast, Duan2014b, braun2016ultrafast, yu2020control}. The photocurrents generated in response to elliptically polarized light can be tuned by shifting the polarization and the angle of incidence. A large body of theoretical literature has also been developed to describe the observed PGE in materials with broken inversion symmetry \cite{Rappe2023, Vorobjev2000, hosur2011circular, Spivak2009, moore2010confinement, morimoto2016topological, xu2021pure}.

Most of the experimental and theoretical work on topological insulators has been devoted to 3D systems. In recent years, more attention has been given to 2D topological insulators as an alternative to 3D materials for their potential applications in spintronics  \cite{wu2023electric, rong2023interaction, hou2020progress, weber20242024, li2015two, zhang2017two}. Two-dimensional TIs are characterized by the quantum spin Hall effect and topologically protected gapless edge states \cite{Kane2005, Kane2005a, Bernevig2006, Qi2006, Konig2007, Zhang2010}. The edge states can be protected against backscattering as a result of the spin-orbit coupling (SOC) and time reversal (TR) symmetry, i.e. in the absence of magnetic field or magnetic impurities. A 2D TI slab is characterized by a band gap \cite{Zhang2010} of about $50$ meV and entangled electrons between its top and bottom surfaces.

Here we study a model of an ultra-thin TI film and examine its out-of-equilibrium charge and spin dynamics in response to electromagnetic radiation. This is motivated both by the general need to understand spin transport in topological 2D materials, and as an exploration of phenomena that could find applications in spintronics. Our model describes an atomically thin film made from a 3D TI material such as $\mathrm{Bi}_{2}\mathrm{Se}_{3}$. The coupling between the top and bottom surfaces of the film opens a band gap in what would otherwise be a topologically protected gapless surface spectrum. Using time-dependent perturbation theory, we construct the detailed-balance equations for the quantum states of electrons in the ultrathin film driven by external radiation. Solving these equations analytically, we determine the radiation-induced charge and spin currents. The net charge and spin currents both vanish unless the symmetry between the top and bottom surfaces is broken -- which generally is the case when a film is grown on a substrate, or when a voltage is applied across the film in a gated device. Projecting on a single surface, the charge currents remain zero but transverse pure spin currents in $s$-wave and $d$-wave channels are generated in response to light with practically any polarization or incidence angle. This effect is a result of the characteristic spin-momentum locking in the spectrum of the TI's electrons and does not require broken mirror symmetries within the TI film as the transverse spin currents are invariant under the in-plane mirror inversions.

The possibility of generating pure spin currents by irradiation of two-dimensional systems has attracted interest in the most recent years \cite{Ruixiang2020, Song2021, Jiang2021, Fang2024}. Our results show that ultrathin TI films are also a useful platform for this effect. While fairly similar spin-current photogalvanic phenomenology appears on the surfaces of 3D TIs \cite{hosur2011circular, Gedik2012, Olbrich2014, Duan2014b, Dantscher2015, kastl2015ultrafast, Okada2016, Plank2016, plank2018infrared}, the 2D geometry offers certain important advantages: (a) no 3D bulk states are involved in optical transitions, (b) there are no intrinsic charge currents and spin accumulations, (c) the system is amenable to control in gated devices, and (d) it may enable inducing and probing the spin currents in strongly correlated states, featuring even electron fractionalization \cite{Nikolic2011a, Nikolic2012a, Nikolic2012, Nikolic2012b, Nikolic2014a}. By applying a gate voltage across the TI film, especially when the film is placed in contact with a metallic substrate (perhaps through an insulating tunneling barrier), one can control the concentrations of electrons in the coupled low energy states at the top and bottom surfaces of the film. This can be used to fine-tune the response of the system to light, and even switch the direction of the induced spin currents. These spin currents and their behavior are also unusual in comparison to the traditionally studied cases. Typical charge and spin photogalvanic effects require a broken mirror symmetry \cite{Ganichev2017}, but not this one. Similarly, circular photogalvanic effect without external bias normally exhibits a sign change of the current upon the reversal of the circular/elliptical polarization handedness \cite{ganichev2003spin}, but this is not a feature of our system's response. Certain methods, which we discuss at the end, are available for the detection of transverse spin currents and their conversion to charge currents.

The rest of this paper is organized as follows. Section \ref{sec:levelII} introduces the model we study and explains its essential features. Section \ref{sec:levelIII} presents the theoretical analysis and results in three stages. We first perturbatively derive the probability rates of light-induced transitions between the electronic states in Section \ref{TDPT}. Then, we construct and solve analytically the detailed balance equations for the occupation numbers of electronic states in Section \ref{DBE}. Finally, we compute the charge and spin currents and present their dependence on the incidence angle, polarization and temperature in Section \ref{RDC}. Implications and possible applications of our findings are discussed in Section \ref{Discussion} and all results are briefly summarized in Section \ref{Conclusions}.

\section{Model}\label{sec:levelII}

The simplest model of an ultrathin film made from a topological band-insulator material such as Bi$_2$Se$_3$ or Bi$_2$Te$_3$ is given by the Hamiltonian:
\begin{equation}\label{H0}
H_0 = \frac{\left(\textbf{p}-\tau^{z}\boldsymbol{\mathcal{A}}\right)^{2}}{2m}+\Delta\tau^{x}-\mu \ .
\end{equation}
In this model, non-interacting electrons have mass $m$, chemical potential $\mu$, and momentum ${\bf p}$ constrained to the two-dimensional $xy$ plane of the film. In addition to spin, the electrons have an internal ``orbital'' degree of freedom which specifies whether they are localized on the top surface $\tau=+1$ or the bottom surface $\tau=-1$ of the TI film. This is a remnant of the 3D TI's topologically protected surface states, now coupled and gapped out by inter-surface tunneling. The Pauli matrices $\tau^{x,y,z}$ govern the inter-surface dynamics; $\tau$ is the eigenvalue of $\tau^z$, and the $\Delta$ term in the Hamiltonian captures the inter-surface tunneling and the ensuing gap-opening. 

The spin-orbit coupling is introduced as a background SU(2) gauge field coupled to electrons:
\begin{equation}\label{SU2}
\boldsymbol{\mathcal{A}}  =mv(\hat{\bf z}\times\boldsymbol{\sigma}) \ , \end{equation}
where $\boldsymbol{\sigma}$ is the vector of spin Pauli matrices. When expanded, the ${\bf p}\boldsymbol{\mathcal{A}} \propto \hat{\bf z}(\boldsymbol{\sigma}\times{\bf p})$ term in the Hamiltonian captures a Rashba-type SOC. We will conveniently represent the components of this gauge field using the antisymmetric Levi-Civita tensor $\epsilon_i^a$ and Einstein's convention for the summation over repeated indices:
\begin{equation}\label{eqn_2.3}
\mathcal{A}_{i}^{\phantom{x}} = A_{i}^{a}\sigma^{a} \quad,\quad
A_{i}^{a} = -mv\epsilon_{i}^{a} \ .
\end{equation}
We will also use the units $\hbar=1$. The eigenvector of $\tau^z$ plays the role of SU(2) charge. The Yang-Mills gauge flux
\begin{equation}
\Phi_i = \epsilon_{ijk} (\partial_j\mathcal{A}_k + i\tau^z \mathcal{A}_j \mathcal{A}_k) = (mv)^2 \delta_{i,3} \sigma^z
\end{equation}
is a spin matrix whose gauge-invariant eigenvalues reveal the presence of an effective ``magnetic field'' that induces topological dynamics, in analogy to the ordinary magnetic field in quantum Hall systems.

With the chemical potential tuned to fall inside the gap, the above Hamiltonian describes a band insulator whose low energy spectrum is given just by the gapped-out surface states of a parent bulk TI. The resulting energy spectrum is
\begin{equation} \label{E0}
E_s({\bf p}) = \frac{p^{2}}{2m}+\frac{mv^{2}}{2} + s\epsilon_p \ ,
\end{equation}
where $s=\pm 1$ is the band index, $p=|{\bf p}|$ and
\begin{equation}
\epsilon_p = \sqrt{(vp)^2+\Delta^{2}} \ .
\end{equation}
The spectrum is degenerate with respect to another quantum number, the spin projection $\sigma=\pm 1$ on the direction perpendicular to ${\bf p}$; this will be referred as to ``helical spin''. The normalized Hamiltonian eigenstates $|{\bf p}s\sigma\rangle$ are superpositions of $\tau_{z}=\pm$ eigenstates $|{\bf p}\sigma\tau_\pm\rangle$ corresponding to the particle residing on the top or bottom surface of the TI film:
\begin{equation}\label{H0states}
|{\bf p}\sigma s\rangle = \frac{\Delta |{\bf p}\sigma\tau_+\rangle + (s\epsilon_p-\sigma vp)|{\bf p}\sigma\tau_-\rangle}{\sqrt{2\epsilon_p(\epsilon_p-\sigma svp)}} \ .
\end{equation}

The charge and spin current operators for a single electron can be defined from the Hamiltonian $H$ as,
\begin{eqnarray}\label{Currents}
j_i &=& i[H,r_i] \\
j_{i}^{a} &=& \frac{1}{2}\left\{ \sigma^{a},j_{i}^{\phantom{x}}\right\} \ . \nonumber
\end{eqnarray}
The lower indices $i,j,\dots$ represent spatial directions, and the upper indices $a,b,\dots$ are spin projections. We do not distinguish between upper and lower space-time indices here as our system is non-relativistic. Substituting the $H=H_0$ from (\ref{H0}) gives:
\begin{eqnarray}\label{eqn_chrgcurrent1}
j_{i}&=&\frac{{p}_{i}-\tau^{z}{\mathcal{A}}_{i}}{m} \\
j_{i}^{a}&=&\frac{1}{2}\left\lbrace \sigma^{a},\frac{{p}_{i}^{\phantom{x}}-\tau^{z}{\mathcal{A}}_{i}}{m}\right\rbrace \ . \nonumber
\end{eqnarray}
The presence of electromagnetic fields is captured by the replacement $p_i \to p_i -\frac{e}{c}A_i$, which includes the gauge potential $A_i$ needed for the gauge invariance.

\section{Photogalvanic Effect}\label{sec:levelIII}

In this section, we derive the charge and spin currents induced in the TI film by a coherent electromagnetic wave with frequency $\omega$. We first use the time-dependent perturbation theory to obtain probability rates for transitions between different electronic states in the TI, and then analyze occupation numbers of single-particle states pushed out of equilibrium by the radiation. This in turn gives us the sought currents as functions of the light polarization and incidence angle.

\subsection{Time-dependent perturbation theory}\label{TDPT}

\begin{figure}
    \centering
    \includegraphics[width=0.6\linewidth]{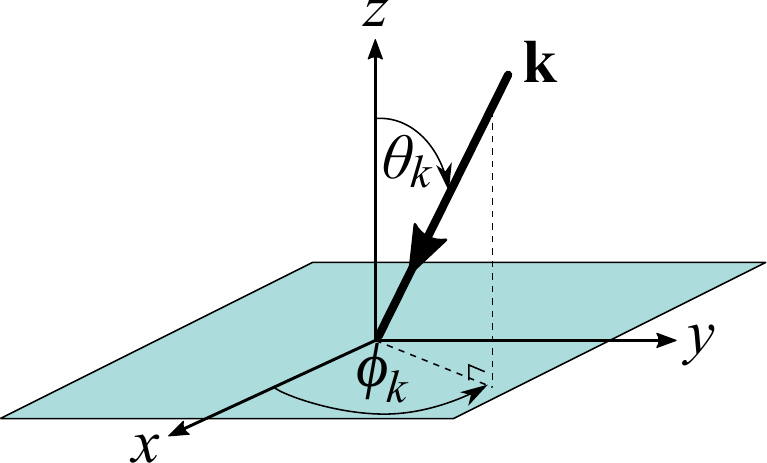}
    \caption{\label{incidence}The geometry of the TI film and incident light with wavevector ${\bf k}$.}
\end{figure}

The electric field of a circularly polarized (CP) light is
\begin{equation} \label{circelectricfield}
{\bf E}(t) = E_{0} \Bigl\lbrack \hat{\bf e}_1 \cos({\bf k}{\bf r}-\omega t)\pm\hat{\bf e}_2 \sin({\bf k}{\bf r}-\omega t) \Bigr\rbrack \ ,
\end{equation}
where $\pm$ indicates the handedness of polarization corresponding to the photons' spin, and ${\bf e}_{1,2}$ are two orthogonal unit vectors perpendicular to the wavevector ${\bf k}=k({\bf e}_1 \times {\bf e}_2)$. Figure (\ref{incidence}) illustrates the incidence of the electromagnetic wave on the TI and the choice of the coordinate system. The propagation wavevector ${\bf k}$ will be taken as
\begin{equation}\label{WaveVect}
\hat{\bf k} = -k \Bigl\lbrack \hat{\bf x}\sin\theta_k\cos\phi_k + \hat{\bf y}\sin\theta_k\sin\phi_k + \hat{\bf z}\cos\theta_k \Bigr\rbrack
\end{equation}
for the incidence angle $\theta_k$. When the CP light is passed through a quarter-wave plate (QWP), it is converted to an elliptically polarized wave characterized by an angle $\varphi$:
\begin{eqnarray}\label{QWPfield}
{\bf E}(t) &=& E_{0} \Bigl\lbrack \hat{\bf e}_1 \cos(\varphi)\cos({\bf k}{\bf r}-\omega t) \\
&& \quad +\hat{\bf e}_2 \sin(\varphi)\sin({\bf k}{\bf r}-\omega t) \Bigr\rbrack \ . \nonumber
\end{eqnarray}
The special cases $\varphi=n\frac{\pi}{2}$ correspond to linear polarization, and $\varphi=(n+\frac{1}{2})\frac{\pi}{2}$ with $n\in\mathbb{Z}$ reproduces circular polarization of either handedness. The subsequent analysis will assume that one symmetry axis (${\bf e}_1$) of the elliptical polarization is kept parallel to the $xy$ plane of electrons. The corresponding scalar $A_0$ and vector ${\bf A}$ potentials will be expressed in the gauge $A_0=0$
\begin{eqnarray}\label{EllPol}
{\bf A}(t) &=& \frac{E_{0}}{k} \Bigl\lbrack \hat{\bf e}_1 \cos(\varphi)\sin({\bf k}{\bf r}-\omega t) \\
&& \quad -\hat{\bf e}_2 \sin(\varphi)\cos({\bf k}{\bf r}-\omega t) \Bigr\rbrack \nonumber
\end{eqnarray}
that reproduces the electromagnetic field via ${\bf E}=-\partial{\bf A}/c\partial t$, ${\bf B}=\boldsymbol{\nabla}\times{\bf A}$. Note that the electromagnetic field modelled here is the one established in the TI film after the reflection and refraction of the external wave take place.

The effect of electromagnetic radiation on electrons is captured by the modified Hamiltonian
\begin{equation} \label{eqn_3.20}
H = \frac{(\textbf{p}-\tau^{z}\boldsymbol{\mathcal{A}}-\frac{e}{c}\textbf{A})^{2}}{2m}-eA_0+\Delta\tau^{x}-\mu \ .
\end{equation}
Representing the SOC with an SU(2) gauge field formally clarifies the manner in which the SOC will enter the perturbation theory. Expanding the quadratic term reveals the light-induced perturbation to (\ref{H0}):
\begin{equation} \label{Hp}
H' = -eA_0 + \frac{e{\bf p}{\bf A}}{mc}-\frac{e\tau^{z}\boldsymbol{\mathcal{A}}{\bf A}}{mc}+\frac{(\frac{e}{c}\textbf{A})^{2})}{2m} \ .
\end{equation}
Considering that the speed of light $c$ is much larger than the typical electron velocity, we will neglect the term proportional to ${\bf A}^2$. However, we will retain the terms linear in ${\bf A}$ because they gain importance from a large SOC.

Using the standard time-dependent perturbation theory, we determine the probability $P_{i\to f}(t)$ of light-induced transitions from the initial ($i$) to the final ($f$) unperturbed stationary states. The unperturbed Hamiltonian is given by (\ref{H0}) and its spectrum (\ref{E0}) will be represented with electronic frequencies $E_i=\hbar\omega_i$ and $E_f=\hbar\omega_f$. At the first order of perturbation theory, one finds
\begin{eqnarray}
&& P_{i\to f} = \frac{1}{\hbar^{2}}\left\vert \int\limits _{0}^{t}dt'\langle f|H'|i\rangle e^{i(\omega_{f}-\omega_{i})t}\right\vert ^{2} \approx \frac{1}{4\hbar^{2}}\sum_{\zeta=\pm1} \\
&& \qquad\quad \times \left\vert \langle f|(iV_{x}+\zeta V_{y})e^{i\zeta{\bf k}{\bf r}}|i\rangle\frac{\sin\left(\frac{1}{2}(\omega_f-\omega_i-\zeta\omega)t\right)}{\frac{1}{2}(\omega_f-\omega_i-\zeta\omega)}\right\vert^{2} \nonumber
\end{eqnarray}
after representing (\ref{Hp}) without its last term as
\begin{equation}
H' \approx V_{x}\cos({\bf k}{\bf r}-\omega t)+V_{y}\sin({\bf k}{\bf r}-\omega t) \ .
\end{equation}
Note that $|i\rangle$ and $|f\rangle$ are normalized to unity for convenience, despite being extended states. The transition probability rates over large time spans assume the Fermi' golden rule form:
\begin{equation}\label{ProbRate1}
\frac{dP_{i\to f}}{dt}=\frac{\pi}{2\hbar^{2}}\sum_{\zeta=\pm1}|\langle f|(iV_{x}+\zeta V_{y})e^{i\zeta{\bf k}{\bf r}}|i\rangle|^{2}\,\delta(\omega_{f}-\omega_{i}-\zeta\omega)
\end{equation}
The elliptical polarization (\ref{EllPol}) yields
\begin{equation}
iV_{x}+\zeta V_{y}=\frac{eE_{0}}{2m\omega}\left(\overleftarrow{{\bf p}}+\overrightarrow{{\bf p}}-2\tau^{z}\boldsymbol{\mathcal{A}}\right)(i\sin\varphi\,\hat{\bf e}_{2}-\zeta\cos\varphi\,\hat{\bf e}_{1}) \nonumber
\end{equation}
where $\overleftarrow{{\bf p}}$ and $\overrightarrow{{\bf p}}$ are momentum operators to be applied to the state on the left and right respectively. The initial and final states are given by (\ref{H0states}) with quantum numbers $|i\rangle = |{\bf p}_i \sigma_i s_i\rangle$ and $|f\rangle = |{\bf p}_f \sigma_f s_f\rangle$. These are extended state, so we must handle the global spatial variations of the electromagnetic field:
\begin{eqnarray}\label{ProbAmpl}
&& \langle f|(iV_{x}+\zeta V_{y})e^{i\zeta{\bf k}{\bf r}}|i\rangle = \frac{eE_{0}}{2m\omega}\,(i\sin\varphi\,\hat{\bf e}_{2}-\zeta\cos\varphi\,\hat{\bf e}_{1}) \nonumber \\
&& \quad\times \Bigl(({\bf p}_{i}+{\bf p}_{f})\langle\sigma_{f}s_{f}|\sigma_{i}s_{i}\rangle+2mv\hat{{\bf z}}\times\langle\sigma_{f}s_{f}|\tau^{z}\boldsymbol{\sigma}|\sigma_{i}s_{i}\rangle\Bigr) \nonumber \\
&& \quad\times \delta_{{\bf p}_{f}-{\bf p}_{i}-\zeta\hbar{\bf k}_{\perp},0} \ .
\end{eqnarray}
Here, $\delta_{{\bf p},0}$ is the Kronecker symbol that implements momentum conservation and facilitates taking the needed squared modulus $|\langle f|\cdots|i \rangle|^2$; it is inherited from the normalization of extended states to unity, in place of the Dirac $\delta$-function
\begin{equation}
\delta_{{\bf p},0} \to \frac{(2\pi)^{2}\hbar^{2}}{A} \delta({\bf p})
\end{equation}
where $A$ is the system area. The $\langle\sigma_{f}s_{f}|\sigma_{i}s_{i}\rangle$ factor in (\ref{ProbAmpl}) is not zero only in intraband transitions $s_i=s_f$, but the electromagnetic radiation has a finite frequency $\omega$ and induces only interband transitions $s_i\neq s_f$ once the momentum conservation is also taken into account. And, photon's momentum is negligible next to the typical electron momenta, so we may approximately neglect the in-plane component ${\bf k}$ of the photon's wavevector and work with ${\bf p}_i={\bf p}_f \equiv {\bf p}$. The remaining matrix element $\langle\sigma_{f}s_{f}|\tau^{z}\boldsymbol{\sigma}|\sigma_{i}s_{i}\rangle$ and other technical details are derived in Appendix \ref{App1}. The final result for the interband transition probability rates in the 2D TI system is
\begin{eqnarray}\label{ProbRate}
&& \frac{dP_{i\to f}}{dt} = \frac{\pi}{A}\left(\pi\frac{eE_{0}v}{\omega}\,\frac{\Delta+\epsilon_p+\sigma_{f}\sigma_{i}\bigl(\Delta-\epsilon_p\bigr)}{2\epsilon_p}\right)^{2} \nonumber \\
&& \quad \times \biggl\lbrace \sin^{2}\varphi\,\cos^{2}\theta_{k}\left\lbrack 1+\sigma_{i}\sigma_{f}\cos\Bigl(2(\phi-\phi_{k})\Bigr)\right\rbrack \nonumber \\
&& \qquad +\cos^{2}\varphi\left\lbrack 1-\sigma_{i}\sigma_{f}\cos\Bigl(2(\phi-\phi_{k})\Bigr)\right\rbrack \biggr\rbrace \nonumber \\
&& \quad \times \sum_{\zeta=\pm1}\delta({\bf p}_{f}-{\bf p}_{i}-\zeta\hbar{\bf k}_{\perp})\,\delta(\omega_{f}-\omega_{i}-\zeta\omega) \ .
\end{eqnarray}

\subsection{Detailed balance equations}\label{DBE}

An electromagnetic wave generally drives an electron gas out of equilibrium by causing transitions between the unperturbed stationary states. The equilibrium occupation of stationary states $|i\rangle$ with energy $E_i$ at temperature $T$ is governed by the Fermi-Dirac distribution
\begin{equation}\label{N0}
n_{i}^{(0)}=\frac{1}{e^{\beta(E_{i}-\mu)}+1} \ ,
\end{equation}
where $\beta=1/k_{\textrm{B}}T$. For simplicity, we will use here a single symbol $i$ to represent all three quantum numbers, momentum ${\bf p}$, spin projection ``helicity'' $\sigma=\pm1$ along the $\hat{{\bf z}}\times\hat{{\bf p}}$ direction, and band index $s=\pm1$. The actual average occupations $n_{i}$ of states $i$ are perturbed out of equilibrium by light according to the detailed balance equations
\begin{eqnarray}
\frac{dn_{i}^{\phantom{x}}}{dt} &=& \sum_{j\neq i}\left\lbrack n_{j}(1-n_{i})\frac{dP_{j\to i}^{\phantom{x}}}{dt}-n_{i}(1-n_{j})\frac{dP_{i\to j}^{\phantom{x}}}{dt}\right\rbrack \nonumber \\
&& -\frac{n_{i}^{\phantom{x}}-n_{i}^{(0)}}{\tau} \ ,
\end{eqnarray}
where $\tau$ is a recombination time associated with spontaneous emission and other processes. The Pauli exclusion is taken into account. Since $dP_{i\to j}/dt=dP_{j\to i}/dt$ according to (\ref{ProbRate}), the detailed balance equations simplify to
\begin{equation}
\frac{dn_{i}^{\phantom{x}}}{dt}=\sum_{j\neq i}(n_{j}-n_{i})\frac{dP_{i\to j}^{\phantom{x}}}{dt}-\frac{n_{i}^{\phantom{x}}-n_{i}^{(0)}}{\tau} \ .
\end{equation}
In a steady state, $dn_i/dt=0$ implies
\begin{equation}\label{DB1}
n_{i}=\frac{n_{i}^{(0)}+\tau\sum_{j\neq i}n_{j}\frac{dP_{i\to j}^{\phantom{x}}}{dt}}{1+\tau\sum_{j\neq i}\frac{dP_{i\to j}^{\phantom{x}}}{dt}} \ .
\end{equation}
This looks like a complicated system of linear equations for $n_{i}$, but the rates $dP_{i\to j}/dt$ are non-zero only if the initial and final states $i,j$ have the same electron momentum when a photon's momentum is negligible, and opposite band index when only inter-band processes occur. Therefore, the system of equations decouples into quadruplets of equations for the states at the same ${\bf p}$ distinguished by $\sigma,s$:
\begin{equation}\label{DB2}
n_{{\bf p},\sigma,s}=\frac{n_{{\bf p},\sigma,s}^{(0)}+A_{{\bf p}}n_{{\bf p},\sigma,-s}+B_{{\bf p}}n_{{\bf p},-\sigma,-s}}{1+A_{{\bf p}}+B_{{\bf p}}} \ .
\end{equation}
The rates $A_{\bf p}$ and $B_{\bf p}$ have absorbed the momentum summations which are implicit in (\ref{DB1}), and this amounts to removing the factors of $\delta_{{\bf p}_j,{\bf p}_i+\zeta\hbar{\bf k}_\perp} = (2\pi)^2 \hbar^2 A^{-1} \delta({\bf p}_j-{\bf p}_i-\zeta\hbar{\bf k}_\perp)$ from (\ref{ProbRate}) after ${\bf k}_\perp$ is neglected,
\begin{eqnarray}\label{ProbRate2}
\frac{dP_{{\bf p},\sigma,s\to {\bf p}',\sigma,-s}}{dt} &=& \frac{A_{\bf p}}{\tau} \frac{(2\pi)^2\hbar^2}{A} \delta({\bf p}'-{\bf p}\pm\hbar{\bf k}_\perp) \\
\frac{dP_{{\bf p},\sigma,s\to {\bf p}',-\sigma,-s}}{dt} &=& \frac{B_{\bf p}}{\tau} \frac{(2\pi)^2\hbar^2}{A} \delta({\bf p}'-{\bf p}\pm\hbar{\bf k}_\perp) \nonumber \ .
\end{eqnarray}
The remaining $\delta$ function of frequency in (\ref{ProbRate}) is also cumbersome to work with. We will replace it with
\begin{eqnarray}
&& \sum_{\zeta=\pm1} \delta(\omega_j-\omega_i-\zeta\omega) \to \\
&& \qquad \to \mathcal{J}(|\omega_j-\omega_i|)
\equiv \frac{e^{-(|\omega_j-\omega_i|-\omega)^{2}/2\delta^{2}}}{\delta\sqrt{2\pi}} \nonumber
\end{eqnarray}
by modelling some realistic frequency bandwidth $\delta$ of the coherent light source. Note that $\omega_{j}-\omega_{i}$ is completely determined by the electron energies at a fixed momentum ${\bf p}$ in the conduction and valence bands,
\begin{equation}\label{DeltaGaussian}
|\omega_{j}-\omega_{i}|=\frac{E_+({\bf p})-E_-({\bf p})}{\hbar}=\frac{2\epsilon_p}{\hbar} \ .
\end{equation}
By using a Gaussian spectrum for light, we will get a smooth distribution of excited particle-hole pairs over the first Brillouin zone. Note that a Lorentzian spectrum is another option, but it would require imposing an ultra-violet cutoff in the momentum integrals over initial and final states. Ultimately, we can extract the dimensionless positive functions
\begin{widetext}
\begin{eqnarray}\label{AB}
A_{{\bf p}} &=& \frac{2\pi\tau\mathcal{I}_0}{c}\left(\frac{ev}{2\hbar\omega}\,\frac{\Delta}{\epsilon_p}\right)^{2}\Bigl\lbrack\sin^{2}\varphi\,\cos^{2}\theta_{k}\,\cos^{2}(\phi-\phi_{k})+\cos^{2}\varphi\,\sin^{2}(\phi-\phi_{k})\Bigr\rbrack \:\mathcal{J}\left(\frac{2\epsilon_p}{\hbar}\right) \\
B_{{\bf p}} &=& \frac{2\pi\tau\mathcal{I}_0}{c}\left(\frac{ev}{2\hbar\omega}\right)^{2}\Bigl\lbrack\sin^{2}\varphi\,\cos^{2}\theta_{k}\,\sin^{2}(\phi-\phi_{k})+\cos^{2}\varphi\,\cos^{2}(\phi-\phi_{k})\Bigr\rbrack \:\mathcal{J}\left(\frac{2\epsilon_p}{\hbar}\right) \nonumber
\end{eqnarray}
\end{widetext}
from (\ref{ProbRate}), (\ref{ProbRate2}) and (\ref{DeltaGaussian}), where $\mathcal{I}_0^{\phantom{x}} = E_0^2 c$ is the energy current density, or intensity, of the incident radiation (up to a constant).

The system of equations (\ref{DB2}) can be represented and solved in matrix form
\begin{equation}\label{DB3}
\mathcal{M}_{\bf p}^{\phantom{0}} N_{\bf p}^{\phantom{0}} = N_{\bf p}^{(0)} \quad\Rightarrow\quad N_{\bf p}^{\phantom{0}} = \mathcal{M}_{\bf p}^{-1} N_{\bf p}^{(0)} \ ,
\end{equation}
where
\begin{equation}
N_{\bf p}^{\phantom{0}} = \left(\begin{array}{c}
n_{{\bf p}++}^{\phantom{x}}\\[0.07in]
n_{{\bf p}+-}^{\phantom{x}}\\[0.07in]
n_{{\bf p}-+}^{\phantom{x}}\\[0.07in]
n_{{\bf p}--}^{\phantom{x}}
\end{array}\right) \quad,\quad
N_{\bf p}^{(0)} = \left(\begin{array}{c}
n_{{\bf p}++}^{0}\\[0.07in]
n_{{\bf p}+-}^{0}\\[0.07in]
n_{{\bf p}-+}^{0}\\[0.07in]
n_{{\bf p}--}^{0}
\end{array}\right)
\end{equation}
and
\begin{equation}
\mathcal{M}_{\bf p} = \left(\begin{array}{cccc}
1\!\!+\!\!A_{{\bf p}}\!\!+\!\!B_{{\bf p}} & -A_{{\bf p}} & 0 & -B_{{\bf p}}\\
-A_{{\bf p}} & 1\!\!+\!\!A_{{\bf p}}\!\!+\!\!B_{{\bf p}} & -B_{{\bf p}} & 0\\
0 & -B_{{\bf p}} & 1\!\!+\!\!A_{{\bf p}}\!\!+\!\!B_{{\bf p}} & -A_{{\bf p}}\\
-B_{{\bf p}} & 0 & -A_{{\bf p}} & 1\!\!+\!\!A_{{\bf p}}\!\!+\!\!B_{{\bf p}}
\end{array}\right)
\end{equation}
The real orthogonal matrix $\mathcal{M}_{\bf p}$ can be inverted analytically.
Since the unperturbed spectrum is degenerate with respect to the helical spin projection $\sigma$, the equilibrium occupation numbers depend only on momentum ${\bf p}$ and the band index $s=\pm 1$, i.e. $n_{{\bf p\sigma s}}^{(0)}=n_{{\bf p}s}^{(0)}$. The solution for the driven occupation numbers reduces to
\begin{equation}\label{DB4}
n_{{\bf p}\sigma s}^{\phantom{x}}-n_{{\bf p}\sigma s}^{(0)}\equiv\delta n_{{\bf p}s}=\frac{A_{{\bf p}}+B_{{\bf p}}}{2A_{{\bf p}}+2B_{{\bf p}}+1}\,s\left(n_{{\bf p},-}^{(0)}-n_{{\bf p},+}^{(0)}\right) \ .
\end{equation}
Note that $n_{{\bf p},-}^{(0)}-n_{{\bf p},+}^{(0)}>0$, and
\begin{equation}
n_{{\bf p},-}^{(0)}-n_{{\bf p},+}^{(0)} \xrightarrow{\mu=0} \tanh\left(\frac{\beta\epsilon_p}{2}\right) \ .
\end{equation}
At zero temperature ($\beta\to\infty$), the maximum excess occupation numbers are $\delta n_{{\bf p}s} \to s/2$ at resonance or when the recombination time $\tau$ is very large ($A_{\bf p}+B_{\bf p}\gg 1$). This can be understood as a result of the balance between absorption and stimulated emission, which dominate the transitions in these regimes; fermionic initial and final states must have the same occupations $n_i=1/2$ in order to support identical rates of photon absorption and stimulated emission.

\subsection{Radiation-driven currents}\label{RDC}

In this section, we derive the steady-state charge $J_\mu^{\phantom x}$ and spin $J_\mu^a$ currents driven by electromagnetic radiation
\begin{eqnarray}\label{Currents2}
{\bf J} &=& \sum_{{\bf p}\sigma s} \delta n_{{\bf p}s} \langle {\bf p}\sigma s| {\bf j} |{\bf p}\sigma s \rangle \\
{\bf J}^{a} &=& \sum_{{\bf p}\sigma s} \delta n_{{\bf p}s} \langle {\bf p}\sigma s| {\bf j}^{a} |{\bf p}\sigma s \rangle \ , \nonumber
\end{eqnarray}
where the stationary states $|{\bf p}\sigma s\rangle$ are given by (\ref{H0states}), the current operators are defined in (\ref{Currents}) and the excess out-of-equilibrium occupation numbers $\delta n_{{\bf p}s}$ are determined from (\ref{DB4}). Writing $\delta n_{{\bf p}s} = s\delta n_{\bf p}$ as suggested in (\ref{DB4}), and evaluating the matrix elements (some details are provided in Appendix \ref{App1}) yields
\begin{eqnarray}
{\bf J} &=& \sum_{s\sigma}\int\frac{d^{2}p}{(2\pi)^{2}}\,s\delta n_{{\bf p}} \frac{{\bf p}+\mathcal{T}_{p\sigma s}\sigma mv\hat{{\bf p}}}{m\hbar^{2}} = 0 \\
{\bf J}^{a} &=& \sum_{s\sigma}\int\frac{d^{2}p}{(2\pi)^{2}}\,s\delta n_{{\bf p}} \frac{\sigma\epsilon^{ab}\hat{p}^{b}{\bf p}+mv\mathcal{T}_{p\sigma s}(\hat{{\bf z}}\times\hat{{\bf x}}^{a})}{m\hbar^{2}} = 0 \nonumber
\end{eqnarray}
for the charge and spin current densities respectively, where
\begin{equation}
\mathcal{T}_{p\sigma s} \equiv \langle{\bf p}\sigma s|\tau^{z}|{\bf p}\sigma s\rangle = \frac{\Delta^{2}-(\epsilon_p-\sigma s vp)^{2}}{2\epsilon_p(\epsilon_p-\sigma svp)} \ .
\end{equation}
The charge currents vanish at least due the time reversal symmetry of $\delta n_{\bf p} = \delta n_{-{\bf p}}$, while the spin currents vanish as as a result of the band index $s=\pm 1$ summation.

Our model so far possesses a perfect symmetry between the top ($\tau^z=+1$) and bottom ($\tau^z=-1$) surfaces of the TI film. However, such a symmetry is generally absent in realistic devices. We can gain insight about the consequences of the top/bottom asymmetry by considering charge and spin currents which are projected to the top and bottom surface of the film. The top/bottom projection operator is
\begin{equation}
\mathcal{P}_{\pm}=\frac{1\pm\tau^{z}}{2} \ .
\end{equation}
Inserting it into the current expressions (\ref{Currents2}) yields projected currents
\begin{eqnarray}\label{Currents3}
\widetilde{{\bf J}}_\pm &=& \sum_{{\bf p}\sigma s} s\delta n_{{\bf p}} \langle {\bf p}\sigma s| \mathcal{P}_{\pm}{\bf j} \mathcal{P}_{\pm} |{\bf p}\sigma s \rangle \\
\widetilde{{\bf J}}^{a}_\pm &=& \sum_{{\bf p}\sigma s} s\delta n_{{\bf p}} \langle {\bf p}\sigma s| \mathcal{P}_{\pm} {\bf j}^{a} \mathcal{P}_{\pm} |{\bf p}\sigma s \rangle \ , \nonumber
\end{eqnarray}
Representing only the non-trivial projected parts $\widetilde{J}=\widetilde{J}_+-\widetilde{J}_-$ of the currents, we find
\begin{eqnarray}
\widetilde{{\bf J}} &=& \sum_{\sigma s}\int\frac{d^{2}p}{(2\pi)^{2}}\,s\delta n_{{\bf p}}^{\phantom{x}}\frac{{\bf p}\mathcal{T}_{p\sigma s}+\sigma mv\hat{{\bf p}}}{m\hbar^{2}}=0 \\
\widetilde{{\bf J}}^{a} &=& \sum_{\sigma s}\int\frac{d^{2}p}{(2\pi)^{2}}\,s\delta n_{{\bf p}}^{\phantom{x}}\frac{\sigma\epsilon^{ab}\hat{p}^{b}{\bf p}\mathcal{T}_{p\sigma s}+mv(\hat{{\bf z}}\times\hat{{\bf x}}^{a})}{m\hbar^{2}}\neq0 \nonumber \ .
\end{eqnarray}
The projected spin currents do not vanish. It is immediately clear that $\widetilde{J}_{i}^{z}=0$ and $\widetilde{J}_{x}^{x}=-\widetilde{J}_{y}^{y}$, but the other components must be examined more carefully because $\delta n_{{\bf p}}$ is not invariant under rotations of ${\bf p}$ about the $z$-axis as a result of the light incidence and polarization. Let us specialize to the in-plane coordinate system in which $\phi_{k}=0$ in (\ref{ProbRate}), i.e. orient the $xy$ coordinate axes to achieve
\begin{eqnarray}
{\bf k} &\xrightarrow{\phi_{k}=0}& -\frac{2\pi}{\lambda}(\hat{{\bf x}}\sin\theta_{k}+\hat{{\bf z}}\cos\theta_{k}) \nonumber \\
{\bf e}_{1} &\xrightarrow{\phi_{k}=0}& -\hat{{\bf y}} \ . \nonumber
\end{eqnarray}
In this coordinate system, we have
\begin{equation}
\delta n_{{\bf p}}\equiv\delta n\left(({\bf p}\hat{{\bf x}})^{2},({\bf p}\hat{{\bf y}})^{2}\right) \ ,
\end{equation}
i.e. changing the sign of the momentum ${\bf p}$ projection along either $x$ or $y$ direction leaves $\delta n_{{\bf p}}$ invariant. Then, $\widetilde{J}_x^x = \widetilde{J}_y^y = 0$, but both transverse spin currents $\widetilde{J}_x^y, \widetilde{J}_y^x$ are finite. We can construct the symmetrized transverse spin currents
\begin{eqnarray}\label{TSC}
\widetilde{J}_s^{\phantom{x}} &=& \widetilde{J}_{y}^{x}-\widetilde{J}_{x}^{y} = \frac{4v}{m\hbar^{2}}\int\frac{d^{2}p}{(2\pi)^{2}}\,\delta n_{{\bf p}}^{\phantom{x}}\,\frac{p^{2}}{\epsilon_{p}} \\
\widetilde{J}_d^{\phantom{x}} &=& \widetilde{J}_{y}^{x}+\widetilde{J}_{x}^{y} = \frac{4v}{m\hbar^{2}}\int\frac{d^{2}p}{(2\pi)^{2}}\,\delta n_{{\bf p}}^{\phantom{x}}\,\frac{p_{y}^{2}-p_{x}^{2}}{\epsilon_{p}} \nonumber
\end{eqnarray}
that exhibit $s$-wave and $d$-wave transformations respectively under rotations about the $z$-axis (i.e. $\widetilde{J}_s$ is invariant, while $\widetilde{J}_d$ changes sign under $\pi/2$ rotations). The projected $s$-wave or ``helical'' spin spin current $\widetilde{J}_s$ exists as a background even in the ground state, but gets enhanced by the absorption of light. The $d$-wave spin current $\widetilde{J}_c$ exists only in the non-equilibrium state since the incoming light violates the rotational symmetry of the $xy$ plane.

The characteristic magnitude $\widetilde{J}_{s,d} \sim J_0$ of all induced currents is controlled by at least two parameters,
\begin{equation}
J_0 = \frac{4\Delta^{3}}{mv^{3}\hbar^{2}} \quad,\quad \alpha = \frac{2\pi\tau}{c\delta}\left(\frac{ev}{2\hbar\omega}\right)^{2}\mathcal{I}_{0} \ .
\end{equation}
Here, $J_0$ is extracted from (\ref{TSC}) by a change of variables that makes the integrals dimensionless; it has the units of number flux in 2D, i.e. velocity per unit area. The parameter $\alpha$ is dimensionless and characterizes the magnitude of $A_{\bf p}$ and $B_{\bf p}$ in (\ref{AB}). Recall that $\tau$ is the electron recombination time, $e$ is electron charge, $\delta$ is the Gaussian frequency bandwidth of the light beam, $\omega$ is the light frequency, and $\mathcal{I}_0^{\phantom{x}}$ is the intensity of light expressed in the energy current density units. We can identify two characteristic regimes according to (\ref{DB4}), weak-intensity $\alpha\ll1$ and strong-intensity $\alpha\gg1$:
\begin{equation}
\widetilde{J}_{s,d}\sim\begin{cases}
\alpha J_{0} & ,\quad\alpha\ll1\\
J_{0} & ,\quad\alpha\gg1
\end{cases} \ .
\end{equation}
The employed perturbation theory holds at least in the weak-intensity regime where the spin current is proportional to the light's intensity. Otherwise, the spin current would saturate if $\alpha$ became large by a very coherent $\delta\to0$ or intense $\mathcal{I}_0\to\infty$ light, or by a very slow spontaneous recombination $\tau\to\infty$.

If the light is very coherent so that $\alpha\gg1$ by $\delta\to0$, then (\ref{DeltaGaussian}) resembles the Dirac $\delta$-function and the excess occupation numbers (\ref{DB4}) behave as:
\begin{eqnarray}
&& \delta n_{{\bf p}s} \propto \frac{A_{\bf p}+B_{\bf p}}{2A_{\bf p}+2B_{\bf p}+1} \\
&& \qquad = \frac{1}{2+\alpha^{-1}F_{\bf p}^{-1}\,e^{\left(2\epsilon_p-\hbar\omega\right)^{2}/2(\hbar\delta)^{2}}} \nonumber \\
&& \qquad \xrightarrow{\delta\to0}
\begin{cases}
  \alpha F_{\bf p} e^{-\left(2\epsilon_p-\hbar\omega\right)^{2}/2(\hbar\delta)^{2}} & ,\quad |2\epsilon_p-\hbar\omega|\gg\hbar\delta \\
  1/2 & ,\quad |2\epsilon_p-\hbar\omega|\ll\hbar\delta
\end{cases} \nonumber
\end{eqnarray}
where $F_{\bf p}\sim \mathcal{O}(1)$ is a dimensionless function. As a result of this, the integrals of (\ref{TSC}) make significant contributions to the spin currents only in a narrow range of momenta $\delta p \propto \delta/\omega$. The $s$-wave current, for example, becomes
\begin{equation}
\lim_{\delta\to0}\widetilde{J}_s \propto \frac{J_0 \delta}{\omega}\, \frac{\left\lbrack \left(\frac{\hbar\omega}{2\Delta}\right)^{2}-1\right\rbrack ^{3/2}}{\sqrt{1-\left(\frac{2\Delta}{\hbar\omega}\right)^{2}}}\,\tanh\left(\frac{\beta\hbar\omega}{4}\right) \ .
\end{equation}
Evidently, the spin current density decreases as the light becomes more coherent, and ultimately vanishes in proportion to $\delta/\omega$. It should be pointed out that this behavior is natural for dispersing non-interacting particles, because energy conservation restricts the particles that absorb a perfectly coherent light ($\delta\to0$) to a ring in momentum space, i.e. a set of measure zero. Without interactions, no other particles can be affected by the incoming coherent light and the number of those on the ring simply does not scale in proportion to the system area. Realistic systems are interacting, however, so the energy absorbed by the particles on the ring can leak to the particles carrying different momenta and produce a macroscopic response to the coherent light that we are not modelling here.

\begin{figure}
\subfigure[{}]{\includegraphics[width=2.8in]{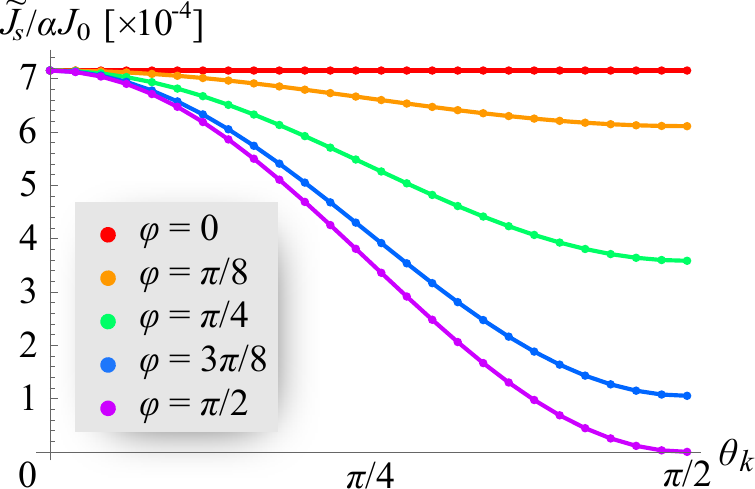}}
\subfigure[{}]{\includegraphics[width=2.8in]{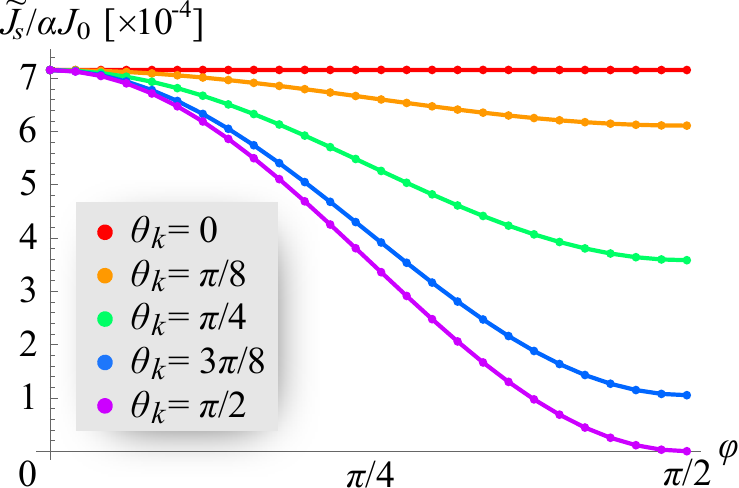}}
\caption{\label{js}The light-induced surface-projected $s$-wave spin current $\widetilde{J}_s$ as a function of (a) the incidence angle $\theta_k$ and (b) polarization angle $\varphi$. The current is expressed in the units of $\alpha J_0$ and calculated at a low temperature $k_{\textrm{B}} T\ll \Delta$, with $\hbar\omega=3\Delta$, $\hbar\delta=0.01\Delta$ and $\alpha\sim0.01$. The difference between $J_s(\theta_k)$ and $J_s(\varphi)$ is very slight for $\alpha\ll 1$.}
\end{figure}

\begin{figure}
\includegraphics[width=2.8in]{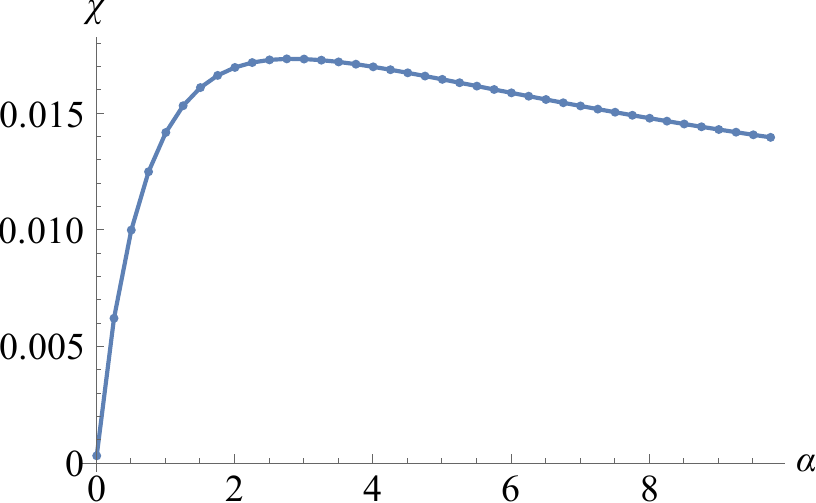}
\caption{\label{js-nf}The deviation from ``flatness'' of the $s$-wave spin current at normal incidence, $\chi=\lbrack\widetilde{J}_s(\varphi=\pi/4)-\widetilde{J}_s(\varphi=0)\rbrack/\widetilde{J}_s(\varphi=0)$. The relevant parameters are $T\to0$, $\hbar\omega=3\Delta$ and $\hbar\delta=0.01\Delta$.}
\end{figure}

The symmetric $s$-wave spin currents projected on a TI surface are shown in Fig.\ref{js} for the weak-intensity regime. If the electromagnetic wave has linear polarization with the electric field oscillating parallel to the electrons' $xy$ plane ($\varphi=0$), then the spin current shows no dependence on the incidence angle $\theta_k$. Recall, however, that our model depicts the electromagnetic field established \emph{inside} the TI film. When the internal electric field ${\bf E}({\bf r},t)$ is constantly parallel to the $xy$ plane, its change with $\theta_k$ is captured merely by the change of the in-plane photon's wavevector ${\bf k}_\perp$, which is negligible in the induced optical transitions. In experimental setups, the amplitude of the external incident field is constant while $\theta_k$ changes, so the internal field we use in calculations may vary as a result of the light reflection and refraction. All other types of polarization are sensitive to the incidence angle via the amplitude of the in-lane electric field. Similarly, the dependence of the spin current of the elliptical polarization angle $\varphi$ turns out to be very slight in general for normal incidence ($\theta_k=0$). The rotational symmetry requires only $\widetilde{J}_s (\varphi=0) = \widetilde{J}_s (\varphi=\pi/2)$ at $\theta_k=0$, because these two cases correspond to two orthogonal linear polarizations, both within the $xy$ plane at normal incidence. Furthermore, the function $\widetilde{J}_s (\varphi)$ is flat at $\theta_k=0$ in the weak-intensity limit, which can be seen after the momentum direction if exactly integrated out for $\alpha\ll 1$ in (\ref{TSC}). Otherwise, some variations in $\widetilde{J}_s (\varphi)$ can be seen for large $\alpha$, although still not very pronounced. If sufficient sensitivity is available, the deviation from flatness in $\widetilde{J}_s (\varphi)$ at normal incidence, depicted in Fig.\ref{js-nf}, can be used to estimate the magnitude of the $\alpha$ parameter.

\begin{figure}
\includegraphics[width=2.8in]{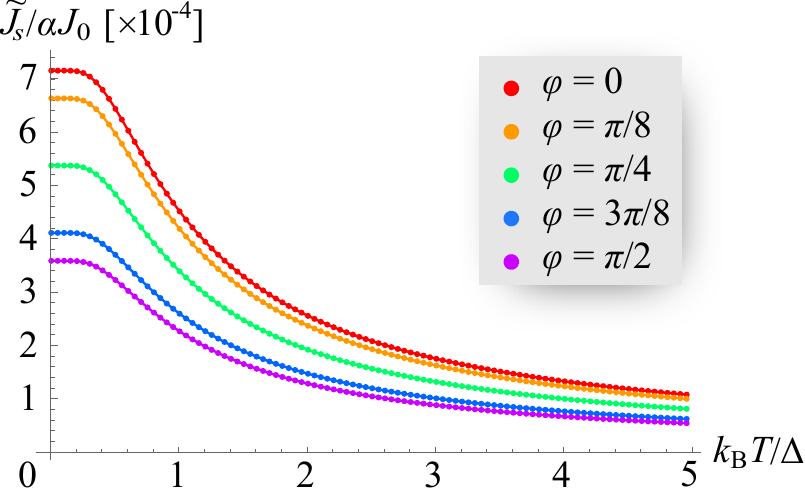}
\caption{\label{js-vs-T}The temperature dependence of the light-induced surface-projected $s$-wave spin current $\widetilde{J}_s$. The current is expressed in the units of $\alpha J_0$ and calculated for the incidence angle $\theta_k=\pi/4$, with $\hbar\omega=3\Delta$, $\hbar\delta=0.01\Delta$ and $\alpha\sim0.01$.}
\end{figure}

The temperature dependence of the $s$-wave spin currents is presented in Fig.\ref{js-vs-T}. The current is maximized at low temperatures when the TI film's conduction band is fully populated with electrons that can freely transition to the empty valence band. When particle-hole excitations become thermally activated, the rate of optical transitions and with it the spin current must decrease. This crossover between the low and high temperature regimes occurs at $k_{\textrm{B}}T \sim \Delta$ as expected.

\begin{figure}
\subfigure[{}]{\includegraphics[width=2.8in]{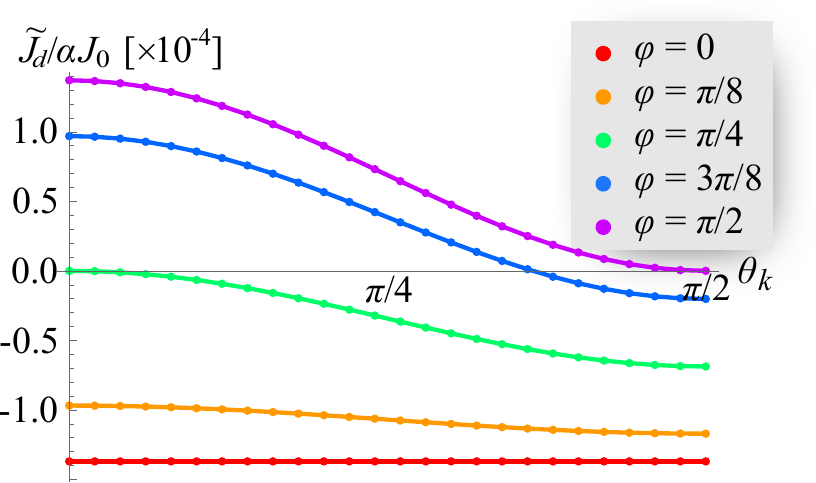}}
\subfigure[{}]{\includegraphics[width=2.8in]{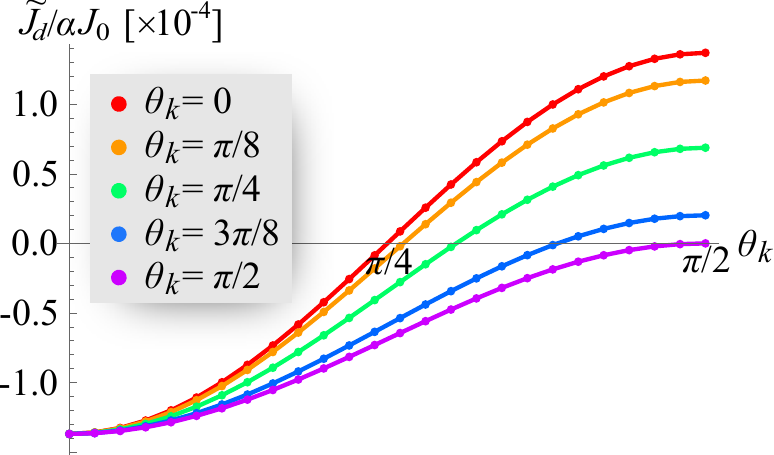}}
\caption{\label{jd}The light-induced surface-projected $d$-wave spin current $\widetilde{J}_d$ as a function of (a) the incidence angle $\theta_k$ and (b) polarization angle $\varphi$ ($0$ is linear in-plane, $\pi/4$ is circular, and $\pi/2$ is linear out-of-plane polarization). The current is expressed in the units of $\alpha J_0$ and calculated at a low temperature $k_{\textrm{B}} T\ll \Delta$, with $\hbar\omega=3\Delta$, $\hbar\delta=0.01\Delta$ and $\alpha\sim0.01$.}
\end{figure}

Fig.\ref{jd} shows the dependence of the $d$-wave spin current projected to a TI surface on the incidence angle $\theta_k$ and polarization $\varphi$. We again find no $\theta_k$ dependence when linearly-polarized electric field lies within the $xy$ plane ($\varphi=0$). However, the dependence of $\varphi$ is now obvious. Note that the $d$-wave current vanishes at normal incidence for the circular polarization $\varphi=\pi/4$; this is a consequence of the rotational symmetry, and the fact that the $d$-wave spin current changes sign under $\pi/2$ rotations about the $z$-axis.

\section{Discussion}\label{Discussion}

Certain features of the calculated currents are a result of various idealizations in our model, most notably the symmetry between the top and bottom surfaces of the TI film. This symmetry is absent in realistic devices and can be further manipulated by gate voltage since the TI film is an insulator. One way to model the asymmetry between the two surfaces is to assign different chemical potentials to them, i.e. introduce a term $\delta H = \delta\mu\,\tau^z$ in the Hamiltonian (\ref{H0}). This produces a charge photocurrent from the obliquely incident light with circular polarization \cite{Dai2012}. The ensuing imbalance in the top/bottom surface electron populations also prevents the exact cancellation between the transverse surface spin currents $\widetilde{J}_\pm^a$, leaving a non-zero net residue $\widetilde{J}_{s}^a = \widetilde{J}_+^a - \widetilde{J}_-^a$. The sign and magnitude of this residue is decided merely by which surface is richer in electrons, so it can be controlled by the gate voltage. In this fashion, a device made from a TI film can realize a spin-current photo-transistor or amplifier. 

A transverse spin current cannot be directly measured. One approach to its detection is to convert it into charge current. A pure transverse spin current consists of counter-propagating electron streams that carry opposite spins. A net charge current arises if some imbalance is introduced between these streams. This can be achieved by applying an external magnetic field parallel to the electrons' plane and the transported spin projection, i.e. by effectively giving different chemical potentials to the electrons in two counter-propagating streams. In practice, one would need to embed magnetic moments in the TI lattice or inside an adjacent ferromagnetic layer of the device heterostructure, then magnetize these moments with external magnetic field or strong current \cite{Kimata2019}.

The last approach is also instrumental for the prospects of transporting angular momentum by photo-excited electrons between two spin reservoirs. The electronic states of the ultrathin TI film are derived from the topologically protected surface states of a parent 3D TI. If such a topological protection is engaged, the suppressed backscattering of spin currents prevents the processes in which angular momentum is drained at some location, transported, and then dumped at another location. Due to the coupling between the surfaces and disorder, spin current backscattering can occur across the area of the TI film, but can also be controlled and reduced by increasing the film thickness \cite{Zhang2010}. Still, the only way to purposefully introduce substantial local spin backscattering for spin transfer between the TI and reservoirs is to break the time-reversal symmetry. This is precisely what the spin reservoirs do locally. Since the methods to control spin reservoirs are also being developed \cite{lou2007electrical, Dankert2015, Kimata2019}, the TI films are promising systems for spintronic applications.

\section{Conclusion}\label{Conclusions}

We analytically derived the charge and spin currents which are excited by electromagnetic radiation in ultrathin topological insulator (TI) films. We utilized for this purpose the first-order time-dependent perturbation theory and detailed balance equations, which offer a glimpse of the steady out-of-equilibrium state driven by light even outside of the linear low-intensity regime. The response we studied amounts to a quadratic spin-current photogalvanic effect. The charge and spin currents vanish in idealized situations with maximum symmetry. An absence of symmetry between the top and bottom surfaces of the TI film results with net transverse spin currents with two distinct components. An $s$-wave component of the spin current, invariant under in-plane rotations, is induced by light of any polarization in excess of the level present in the equilibrium ground state. An additional $d$-wave component, which changes sign under $\pi/2$ rotations, exists only out of equilibrium. Both spin currents are invariant under in-plane mirror inversions, but their sign and magnitude can be externally manipulated by a voltage applied across the TI film. We also discussed the means to convert such spin currents into charge currents that would facilitate detection.

\section{Acknowledgements}\label{secAck}

We are grateful to Patrick Vora and Mingzhen Tian for helpful discussions. This research was carried out and supported at the Quantum Science and Engineering Center of George Mason University (GMU).

\appendix
\section{Optical transition matrix elements}\label{App1}

Here we calculate the interband matrix elements $\langle\sigma_{f}s_{f}|\tau^{z}\boldsymbol{\sigma}|\sigma_{i}s_{i}\rangle$ of the perturbation Hamiltonian in the basis of stationary states (\ref{H0states}), and complete the derivation of (\ref{ProbRate}) from (\ref{ProbRate1}). Since the spatial degrees of freedom have already been integrated out, the initial $|\sigma_{i}s_{i}\rangle$ and final $|\sigma_{f}s_{f}\rangle$ states are represented at a fixed momentum ${\bf p}$ in a 4-dimensional vector space as suggested by (\ref{H0states}). We can separate the spin and ``orbital'' degrees of freedom as
\begin{equation}
\langle\sigma_{f}s_{f}|\tau^{z}\boldsymbol{\sigma}|\sigma_{i}s_{i}\rangle \xrightarrow[s_i\neq s_f]{\textrm{interband}} T_{fi} \langle\sigma_{f}|\boldsymbol{\sigma}|\sigma_{i}\rangle \ .
\end{equation}
for interband processes $s_{i}\neq s_{f}$, where
\begin{eqnarray}
&& T_{fi} = \frac{\Delta^{2}+(\epsilon_p\pm\sigma_{f}vp)(\epsilon_p\mp\sigma_{i}vp)}{2\epsilon_p\sqrt{\epsilon_p\pm\sigma_{f}vp}\sqrt{\epsilon_p\mp\sigma_{i}vp}} =
\begin{cases}
  \frac{\Delta}{\epsilon_p} & ,\quad \sigma_{i}=\sigma_{f} \\
  1 &,\quad \sigma_{i}\neq\sigma_{f}
\end{cases} \nonumber
\end{eqnarray}
This can be summarized as
\begin{equation}\label{ME_tau}
T_{fi} = \frac{\Delta+\epsilon_p+\sigma_{f}\sigma_{i}\bigl(\Delta-\epsilon_p\bigr)}{2\epsilon_p} \ .
\end{equation}

The spin matrix elements $\langle\sigma_{f}|\boldsymbol{\sigma}|\sigma_{i}\rangle$ can be calculated using the spinor representation of the stationary states. Since the stationary states are eigenstates of $(\hat{{\bf p}}\times\hat{{\bf z}})\boldsymbol{\sigma}$ in (\ref{H0}), and $\hat{{\bf p}}=\hat{{\bf x}}\cos\phi+\hat{{\bf y}}\sin\phi$ lies in the $xy$-plane, the electron spin is oriented in the $\hat{{\bf n}}=\hat{{\bf p}}\times\hat{{\bf z}}=\hat{{\bf x}}\sin\phi-\hat{{\bf y}}\cos\phi$ direction. The corresponding spinors in the standard representation that diagonalizes $\sigma^z$ are:
\begin{equation}
|\sigma_+\rangle=\frac{1}{\sqrt{2}}\left(\begin{array}{c}
1\\
-ie^{i\phi}
\end{array}\right)\quad,\quad|\sigma_-\rangle=\frac{1}{\sqrt{2}}\left(\begin{array}{c}
ie^{-i\phi}\\
-1
\end{array}\right) \ .
\end{equation}
We readily find
\begin{eqnarray}
\langle\sigma_+|\boldsymbol{\sigma}|\sigma_+\rangle = \hat{{\bf n}} \quad &,& \quad
\langle\sigma_-|\boldsymbol{\sigma}|\sigma_-\rangle = -\hat{{\bf n}} \nonumber \\
\langle\sigma_+|\boldsymbol{\sigma}|\sigma_-\rangle &=& e^{-i\phi}(-\hat{{\bf p}}+i\hat{{\bf z}}) \nonumber
\end{eqnarray}
when the momentum transfer from photons to electrons is neglected, so that the initial and final electron states have approximately the same momentum ${\bf p}_i\approx{\bf p}_f$ that points in the direction given by $\phi$ within the $xy$ plane. This can be summarized with:
\begin{eqnarray}\label{ME_sigma}
&& \langle\sigma_{f}|\boldsymbol{\sigma}|\sigma_{i}\rangle = e^{i(\sigma_{i}-\sigma_{f})\phi/2}\biggl\lbrack -\hat{{\bf x}}\cos\left(\phi+\frac{\sigma_{f}+\sigma_{i}}{2}\frac{\pi}{2}\right) \phantom{MMM} \\
&& \quad -\hat{{\bf y}}\sin\left(\phi+\frac{\sigma_{f}+\sigma_{i}}{2}\frac{\pi}{2}\right)+i\hat{{\bf z}}\sin\left(\frac{\sigma_{f}-\sigma_{i}}{2}\frac{\pi}{2}\right)\biggr\rbrack \nonumber
\end{eqnarray}

Having the matrix elements, we can proceed with the calculation of the transition probability rates. We need to calculate the squared modulus of (\ref{ProbAmpl}) for interband transitions:
\begin{eqnarray}\label{VMod2}
&& |\langle f|(iV_{x}+\zeta V_{y})e^{i\zeta{\bf k}{\bf r}}|i\rangle|^2 \propto (2mv)^{2}\left\vert T_{fi}\right\vert^{2} \phantom{MMM} \\
&& \qquad\qquad\qquad \times \left\vert (-i\sin\varphi\,\hat{\bf e}_{2}+\zeta\cos\varphi\,\hat{\bf e}_{1})\hat{\boldsymbol{\xi}}_{\sigma_{i}\sigma_{f},{\bf p}}\right\vert ^{2} \ , \nonumber
\end{eqnarray}
where we have defined
\begin{equation}
\hat{{\bf z}}\times\langle\sigma_{f}|\boldsymbol{\sigma}|\sigma_{i}\rangle = \pm e^{i(\sigma_{i}-\sigma_{f})\phi/2}\,\hat{\boldsymbol{\xi}}_{\sigma_{i}\sigma_{f},{\bf p}} \ .
\end{equation}
Using (\ref{ME_sigma}), we find
\begin{equation}\label{Xi}
\hat{\boldsymbol{\xi}}_{\sigma_{i}\sigma_{f},{\bf p}}=\begin{cases}
\hat{{\bf p}} & ,\quad\sigma_{i}\sigma_{f}=+1\\
\hat{{\bf p}}\times\hat{{\bf z}} & ,\quad\sigma_{i}\sigma_{f}=-1
\end{cases} \ .
\end{equation}
To make further progress, we define the polarization unit-vectors
\begin{eqnarray}\label{E1E2}
{\bf e}_{1} &=& \hat{{\bf x}}\sin\phi_{k}-\hat{{\bf y}}\cos\phi_{k} \\
{\bf e}_{2} &=& -\hat{{\bf x}}\cos\theta_{k}\cos\phi_{k}-\hat{{\bf y}}\cos\theta_{k}\sin\phi_{k}+\hat{{\bf z}}\sin\theta_{k} \nonumber \ ,
\end{eqnarray}
such that ${\bf e}_1$ is parallel to the $xy$ plane, ${\bf e}_1 \cdot {\bf e}_2 = 1$ and ${\bf e}_1 \times {\bf e}_2 = {\bf k}/k$ reproduces the wavevector (\ref{WaveVect}) of the electromagnetic wave. Substituting (\ref{Xi}) and (\ref{E1E2}) into (\ref{VMod2}) yields after some algebra
\begin{eqnarray}\label{VMod2}
&& |\langle f|(iV_{x}+\zeta V_{y})e^{i\zeta{\bf k}{\bf r}}|i\rangle|^2 \propto  \frac{(2mv)^{2}}{2}\left\vert T_{fi}\right\vert ^{2} \\
&& \qquad \times \biggl\lbrace \sin^{2}\varphi\,\cos^{2}\theta_{k}\left\lbrack 1+\sigma_{i}\sigma_{f}\cos\Bigl(2(\phi-\phi_{k})\Bigr)\right\rbrack \nonumber \\
&& \qquad~~ +\cos^{2}\varphi\left\lbrack 1-\sigma_{i}\sigma_{f}\cos\Bigl(2(\phi-\phi_{k})\Bigr)\right\rbrack \biggr\rbrace  \ . \nonumber
\end{eqnarray}
Together with (\ref{ME_tau}), this completes the derivation of (\ref{ProbRate}).


%

\end{document}